\documentclass[conference]{IEEEtran}
\usepackage{myStyleIEEE}
\usepackage{ushort}
\IEEEoverridecommandlockouts

\usepackage{siunitx}
\AtBeginDocument{\DeclareSIUnit{\MWh}{MWh}}
\AtBeginDocument{\DeclareSIUnit{\kWh}{kWh}}
\AtBeginDocument{\DeclareSIUnit{\voltampere}{VA}}

\usepackage{amsmath}
\usepackage{mathtools}
\usepackage{isomath}

\usepackage{eurosym}
\DeclareRobustCommand{\officialeuro}{%
  \ifmmode\expandafter\text\fi
  {\fontencoding{U}\fontfamily{eurosym}\selectfont e}}

\usepackage{acronym}
\usepackage[]{optidef}
\usepgfplotslibrary{groupplots}
\definecolor{darkgreen}{rgb}{0.0, 0.5, 0.0}

\begin{document}

\acrodef{RES}{Renewable Energy Source}
\acrodef{SG}{Synchronous Generator}
\acrodef{VSC}{Voltage Source Converter}
\acrodef{GFM}{Grid-forming}
\acrodef{DAE}{Differential Algebraic Equation}

\title{Comparison of Current-Limiting Strategies for Grid-Forming Converters Under Grid Faults}
\renewcommand{\theenumi}{\alph{enumi}}

\newcommand{\maitraya}[1]{\textcolor{magenta}{$\xrightarrow[]{\text{M}}$ #1}}
\newcommand{\xiuqiang}[1]{\textcolor{blue}{$\xrightarrow[]{\text{X}}$ #1}}
\newcommand{\ramona}[1]{\textcolor{red}{$\xrightarrow[]{\text{R}}$ #1}}

 \author{
 \IEEEauthorblockN{Maitraya Avadhut Desai\IEEEauthorrefmark{1}, Ramona Stoll\IEEEauthorrefmark{2}, Xiuqiang He\IEEEauthorrefmark{3}, Florian Dörfler\IEEEauthorrefmark{2}, Gabriela Hug\IEEEauthorrefmark{1}}%
 \IEEEauthorblockA{\IEEEauthorrefmark{1} Power Systems Laboratory, ETH Zurich, Switzerland, \IEEEauthorrefmark{2}Automatic Control Laboratory, ETH Zurich, Switzerland,} 
\IEEEauthorblockA{\IEEEauthorrefmark{3} Department of Automation, Tsinghua University, China} 
{Emails: \{desai, hug\}@eeh.ee.ethz.ch, \{rstoll, dorfler\}@ethz.ch, hxq19@tsinghua.org.cn}
\thanks{Research supported by NCCR Automation, a National Centre of Competence in Research, funded by the Swiss National Science Foundation (grant number 51NF40\_180545).}
}

\maketitle
\IEEEpeerreviewmaketitle

\begin{abstract}

This paper investigates the performance of three state-of-the-art current-limiting strategies for grid-forming (GFM) converters under grid faults. These highly effective strategies include virtual admittance control, virtual admittance control enhanced with virtual power feedback, and the recently developed cross-forming control. The performance comparison addresses two crucial requirements for GFM converters under grid faults, namely transient stability and reactive current injection. The comparison indicates that the virtual admittance control enhanced with virtual power feedback shows the most promising transient stability performance, while the cross-forming control exhibits superior reactive current injection performance. Thus, individual benefits of each strategy can be obtained by properly combining them.
\end{abstract}


\section{Introduction} \label{sec:intro}

Grid-forming (GFM) converters enable the smooth transition from synchronous machine-dominated to converter-dominated power systems \cite{Milano-review}. GFM converters are controlled as voltage sources behind an impedance during stable operation \cite{lasseter2019grid}. These converters contribute to the strength of the grid by autonomously regulating voltage and frequency. However, during large disturbances, limiting the output current of the converter is critical to prevent potential overcurrent damage.

Under grid faults, successful fault ride-through (FRT) for the converters is characterized by maintaining transient stability and providing ancillary services such as fault reactive current and phase-jump active power. Recent GFM specifications from Great Britain \cite{gb2024}, Australia \cite{aemo2023voluntary} and Europe \cite{entso2024} list these ancillary services as mandatory requirements. Thus, even when the converter current reaches the limit, it must maintain synchronism and provide the mandatory services \cite{bahrani2024grid,ghimire2024functional}. Meeting all of the above criteria requires a refined current-limiting strategy for GFM converters \cite{baeckeland2024overcurrent,ordono2024current}, for which various techniques have recently emerged.

The adaptive/threshold virtual impedance type current-limiting strategy uses current feedback to increase the virtual impedance magnitude adaptively. Thus, the voltage reference is reduced and the current is kept within bounds \cite{paquette2015virtual,qoria2020current}. Although this method maintains synchronism, it underutilizes the overcurrent capability and involves cumbersome parameter tuning. Switching to a grid-following control during the fault has also been proposed\cite{li2023transient,wang2023transient}. Although this method effectively limits the current, it is unable to provide voltage-forming behavior and meet the grid codes. As an alternative, a current limiter cascaded with a virtual admittance can be used as an effective strategy, as the admittance acts as a proportional regulator \cite{fan2022equivalent,zhang2023current,saffar2023impacts}. Thus, the wind-up issues inherent in classical proportional integral (PI) or proportional resonant (PR) regulators are avoided. Moreover, this strategy fully utilizes the overcurrent capability and is easy to tune. A virtual power enhancement has been proposed for this strategy, which uses the unsaturated current for the power feedback to ensure transient stability \cite{kkuni2024effects,virtpower_laba_2023}. Thus, this strategy, with the proposed enhancements, is able to preserve synchronism and provide ancillary services. Finally, the recently proposed cross-forming control is also an effective current-limiting strategy for grid-forming converters under grid faults \cite{He_crossform_2025}. This strategy ensures synchronization and the provision of ancillary services by preserving voltage angle-forming characteristics while also forming the current magnitude in order to respect the current limit. All of these current-limiting strategies enhance the stability of the power system; however, their relative merits are unclear, and they must be validated and compared against each other for different faults in the grid.

To this end, this paper aims to analyze the differences between the current-limiting strategies and compare them against each other for different grid faults. Furthermore, we focus on both transient stability and the provision of ancillary services mandated by the grid codes. Specifically, this paper has two substantial contributions. First, using the equivalent circuit of the GFM converter with various current-limiting strategies, we analyze the difference in the resulting transient stability characteristics. Second, we compare the strategies for the provision of reactive current during a grid voltage dip in terms of response time and magnitude. Furthermore, we provide the results of detailed electromagnetic transient (EMT) simulations considering all combinations of grid voltage dips and phase jumps to validate the findings.

\section{Grid-Forming Control Configuration} \label{sec:GFMcontrol}

\subsection{Generic GFM Control Structure}

\begin{figure}
  \begin{center}
  \includegraphics{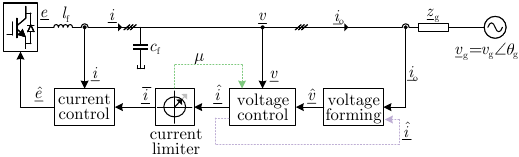}
   \vspace{-0.4cm}
  \caption{Typical control diagram of a grid-connected grid-forming converter. The purple and green dotted lines represent the modified connections for the virtual power and cross-forming methods, respectively.}
   \vspace{-0.4cm}
  \label{fig:generic-control-diagram}
  \end{center}
\end{figure}
In this paper, a grid-connected GFM converter is considered as shown in Fig.~\ref{fig:generic-control-diagram}. A cascade control structure is used where the outer voltage-forming loop generates the reference voltage. The inner controllers track the reference provided and a circular current limiter is placed between the inner controllers to modify the current reference to achieve the current limitation. A balanced three-phase system is assumed and the modeling and analysis is done using complex vector coordinates, e.g., $\pha v \coloneqq v_{\alpha} + j v_{\beta}=v \angle \theta$.

The outer voltage-forming loop can be implemented using any of various grid-forming methods present in the literature. In this paper, we focus on the most commonly used droop control and virtual synchronous machine (VSM) control. The linear droop control is given as 
\begin{subequations}
    \begin{align}
        \dot {\hat\theta} = \hat \omega &= \omega_0 + m_p (p^{\star} - p), \\
        \hat v &= v^{\star} + m_q (q^{\star} - q)\label{eq:reactivedroop},
    \end{align}
    \label{eq:droop}
\end{subequations}
where $m_p \in \mathbb{R}_{\geq 0}$ and $m_q  \in \mathbb{R}_{\geq 0}$ denote the power droop gains, $p^{\star}$, $q^{\star}$ and $v^{\star}$ denote the respective power and voltage setpoints and the nominal frequency is given as $\omega_0$. The VSM maintains the same reactive power droop control as \eqref{eq:reactivedroop}, however, it mimics the dynamics of the synchronous machine with the help of inertia and damping characteristics. It is given as
\begin{subequations}
    \begin{align}
        \dot {\hat \theta} &= \hat \omega, \\
        T_{J} \dot {\hat \omega} &= -D(\hat \omega - \omega_0) + (p^{\star} - p), 
    \end{align}
    \label{eq:vsm}
\end{subequations}
with virtual inertia time constant $T_{J}$ and damping gain $D$.

The inner controllers can be realized in the $\alpha\beta$ frame using PR control \cite{desai2023saturation} or in the $dq$ frame using PI control \cite{Subotic-dVOC}. Between the two inner controllers, the circular current limiter reduces the current reference magnitude only when $\pha{\hat i}$ exceeds $i_{\lim}$, keeping the phase angle unchanged. The output, denoted as $\pha{\bar i}$, is given by 
\begin{gather}
    \label{eq:CircCurrLim}
    \pha{\bar i} =
    \begin{cases}
        \pha{\hat i} & \text{if } \lvert\pha{\hat i}\rvert \leq i_{\lim},\\
          \frac{i_{\lim}}{\lvert\pha{\hat i}\rvert}\pha{\hat i} & \text{if } \lvert\pha{\hat i}\rvert > i_{\lim}.
    \end{cases}
\end{gather}
The degree of saturation (DoS) of the current limiter can be defined as $\mu\coloneqq\pha{\bar i}/\pha{\hat i}\in (0,1]$.

\subsection{Current-Limiting Strategies and Equivalent Circuits}
The dynamics of the inner current control loop and the electromagnetic dynamics of the circuit are generally ignored in studies involving current-limiting strategies \cite{huang2019transient}. Therefore, the differences in the current-limiting strategies lie in the outer voltage-forming control loop and the inner voltage-control loop. 

\begin{figure}
  \begin{center}
  \includegraphics{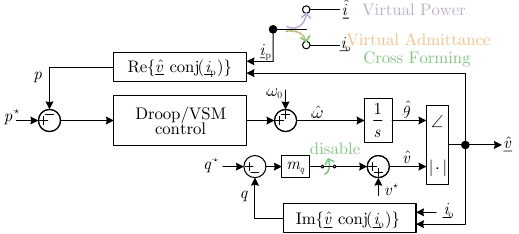}
  \vspace{-0.6cm}
  \caption{Outer voltage-forming control loop for the three current-limiting strategies. The modifications for the virtual admittance, virtual power and cross-forming strategies are shown in orange, purple and green, respectively.}
  \vspace{-0.2cm}
  \label{fig:outer_loop}
  \end{center}
\end{figure}

\begin{figure}
  \begin{center}
  \includegraphics{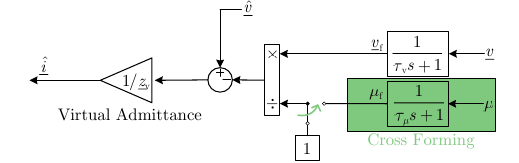}
   \vspace{-0.7cm}
  \caption{Inner voltage control using the virtual admittance strategy and using the cross-forming strategy.}
       \vspace{-0.4cm}
  \label{fig:inner_volt_control}
  \end{center}
\end{figure}

\subsubsection{Virtual Admittance Control}

\begin{figure}
  \begin{center}
  \vspace{-0.3cm}
  \includegraphics{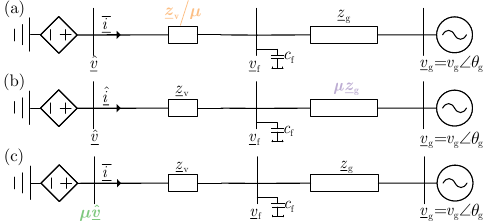}
  \caption{Equivalent circuit obtained using the (a) virtual admittance control (b) virtual power enhancement (c) cross-forming strategy.}
  \label{fig:eqtckt}
  \end{center}
\end{figure}

The virtual admittance current-limiting strategy maintains the standard form for the outer voltage-forming control loop. As seen in Fig.~\ref{fig:outer_loop}, the active power calculation performed before the implementation of the droop or VSM control is given by $p={\rm{Re}} \{\hat{\pha v}\,{\rm{conj}} ({\pha{i}}_{\rm o})\}$.

To avoid wind-up issues during current limitation caused by the inner PR voltage controllers, this strategy uses a complex gain proportional regulator to generate the current reference. This acts as a virtual admittance, particularly in the case of current saturation, as shown in the control diagram in Fig.~\ref{fig:inner_volt_control}. The current reference is given as \cite{fan2022equivalent,zhang2023current,saffar2023impacts}
\begin{equation}
\label{VirtualAdmittance}
    \pha{\hat i} = \frac{1}{\pha{z}_{\rm v}}\left({\pha{\hat{v}}}- {\pha{v}_{\rm f}}\right),
\end{equation}
where ${\pha{v}_{\rm f}}$ is the filtered voltage and $\pha{z}_{\rm v}=r_{\rm v}+jx_{\rm v}$ acts as the virtual impedance behind the capacitor terminal.

When the current is limited, the DoS can be used along with \eqref{VirtualAdmittance} to obtain the output current as
\begin{equation}
    \pha{i} = \pha{\bar i} = \mu \pha{\hat i} = \mu \frac{1}{\pha{z}_{\rm v}}\left({\pha{\hat{v}}}- {\pha{v}_{\rm f}}\right).
    \label{PIopcurropsat}
\end{equation}
Using \eqref{PIopcurropsat}, the equivalent circuit of the grid-connected converter, with the virtual admittance strategy, can be obtained as shown in Fig.~\ref{fig:eqtckt}(a) \cite{zhang2023current}.

\subsubsection{Virtual Power Enhancement}
The virtual power method is an enhancement of the virtual admittance current-limiting strategy. As seen in Fig.~\ref{fig:outer_loop}, the active power calculation is performed in the implementation of the droop or VSM control by applying the unsaturated current, i.e., ${\pha{i}}_{\rm p}={\pha{\hat{i}}}$. However, the inner voltage control remains the same as shown in Fig.~\ref{fig:inner_volt_control}.

To obtain the equivalent circuit of the virtual power enhancement method during current limitation, we note that the virtual unsaturated current $\pha{\hat i}$ flows from the internal voltage while the saturated current $\pha{\bar i}=\mu \pha{\hat i}$ flows into the grid. This can also be simplified by considering the grid impedance to be $\pha{z}'_{\rm g}=\mu \pha{z}_{\rm g}$ as shown in Fig.~\ref{fig:eqtckt}(b) \cite{kkuni2024effects}. 

\subsubsection{Cross-Forming Strategy}
The cross-forming current-limiting strategy uses the output current measurement for the active power calculation in the outer loop, i.e., $\pha{i}_{\rm p}=\pha{i}_{\rm o}$ as shown in Fig.~\ref{fig:outer_loop}. Furthermore, the voltage-magnitude droop control is also disabled as it is not independently controlled during current saturation. The virtual admittance voltage controller is modified as the feedback voltage is scaled by the filtered DoS $\mu_{\rm f}$ and is represented as
\begin{equation}
        \pha{\hat i} = \frac{1}{\pha{z}_{\rm v}}\Bigl({\pha{\hat{v}}}- \frac{{\pha{v}}_{\rm f}}{\mu_{\rm f}}\Bigr).
        \label{PIopvsatMod_stable}
\end{equation}
The internal virtual voltage is defined as $ \hat{\pha v}_{\mu} \coloneqq \mu_{\rm f} \hat{\pha v}$. The converter output current can therefore be written as 
\begin{equation}
    \pha{i} = \pha{\bar i} \approx \mu_{\rm f} \pha{\hat i} =  \frac{1}{\pha{z}_{\rm v}}\left({\pha{\hat{v}}_{\mu}}-{\pha{v}_{\rm f}}\right),
    \label{eq:PIopcurropsat_stable}
\end{equation}
while ignoring the dynamics of the low-pass filter. Thus, using \eqref{eq:PIopcurropsat_stable}, the equivalent circuit of the grid-connected cross-forming converter can be obtained as shown in Fig.~\ref{fig:eqtckt}(c) \cite{He_crossform_2025}.

\section{Transient Stability Performance}

\subsection{Comparative Analysis}

For different current-limiting strategies, their transient stability can be analyzed by studying the power-angle relationship in the equivalent circuits of the grid-connected converters. The following stability analysis assumes the decoupled active power--frequency and reactive power--voltage relationship and considers only the inductive parts of the impedances. This is valid since the focus of this study is the application of converters in large power systems that have a high $x/r$ ratio. Moreover, for ease of analysis, we also ignore the large impedance of the filter capacitor.
When operating the GFM converter using the VSM control mode \eqref{eq:vsm}, transient stability or synchronism with the grid is analyzed in the period after the fault has been cleared. On the other hand, for the GFM converter with (inertia-less) droop control mode \eqref{eq:droop}, synchronism with the grid is analyzed during the fault, since the inertia-less droop control acts fast and can stabilize or destabilize within the duration of the fault \cite{pan2020transient}.

\subsubsection{Virtual Admittance (VA) Control} The power-angle relationship relies on the voltage reference $\pha{\hat v}$ and the measured current ${\pha{i}}_{\rm o}$, which is approximately equal to the saturated current reference $\pha{\bar i}$. Using the equivalent circuit shown in Fig.~\ref{fig:eqtckt}(a), the power-angle relationship can be expressed as
\begin{equation} 
 p_{\mathrm{VA}} = \mathrm{Re} \left( \pha{\hat v} \cdot \mathrm{conj}\left( {\pha{i}}_{\rm o} \right) \right) = \frac{{v}_{\rm g} \cdot \hat{v}}{{x}_{\rm v}/\mu + {x}_{\rm g}} \sin\delta, 
 \label{eq:p_VA}
\end{equation}
where $\delta=\hat{\theta}-\theta_{\rm g}$. As $\mu < 1$, the current-dependent equivalent impedance increases, resulting in a reduction in the magnitude of the power-angle curve. Furthermore, $\mu$ can be obtained as a function of the other variables using the relation
\begin{equation} 
 i_{\rm lim} = \bigg|{\biggl(\frac{\lvert \pha{\hat v}\rvert \angle \hat{\theta} - v_{\mathrm{g}} \angle \theta_{\mathrm{g}}}{j x_{\mathrm{v}}/\mu + j x_{\mathrm{g}}} \biggr)}\bigg|. 
 \label{eq:mu_p_VA}
\end{equation}
Since $\mu$ is related to the angle $\delta$ by \eqref{eq:mu_p_VA}, the power expression \eqref{eq:p_VA} is not sinusoidal with respect to $\delta$ but is a different nonlinear function of it.

\begin{figure}
  \begin{center}
  \includegraphics{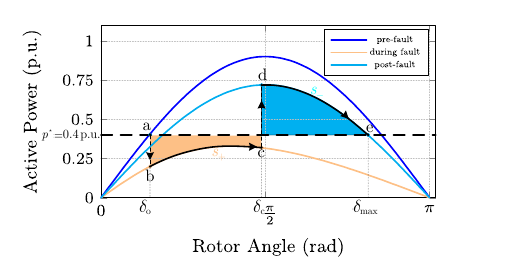}
   \vspace{-0.6cm}
  \caption{Power-angle curves for the virtual admittance current-limiting strategy considering the VSM control operation mode. }
   \vspace{-0.4cm}
  \label{fig:powerangle_VA_VSM}
  \end{center}
\end{figure}

\begin{figure}
  \begin{center}
  \includegraphics{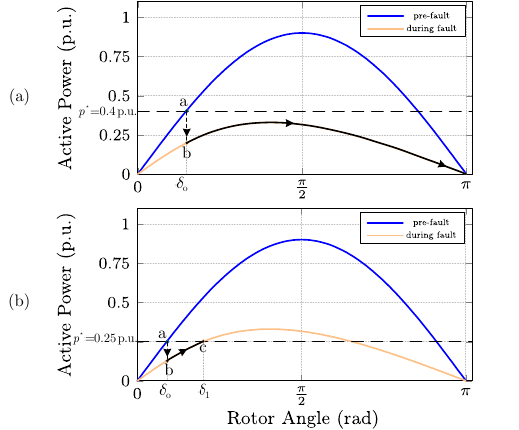}
   \vspace{-0.6cm}
  \caption{Power-angle curves for the virtual admittance strategy considering the droop control operation mode for (a) $p^\star=0.4\,\rm{p.u.}$ and (b) $p^\star=0.25\,\rm{p.u.}$. }
   \vspace{-0.47cm}
  \label{fig:powerangle_VA_droop}
  \end{center}
\end{figure}

To illustrate, we consider a grid-connected GFM converter with a fault such that the grid voltage magnitude drops from $v_{\rm g}=1\,\rm{p.u.}$ to $v_{\rm g}=0.3\,\rm{p.u.}$. A standard set of parameters is chosen to plot the power-angle curves shown in Figs.~\ref{fig:powerangle_VA_VSM} and~\ref{fig:powerangle_VA_droop} for the VSM and droop control modes, respectively. 

For the VSM control operation mode, the power setpoint is chosen as $p^{\star}=0.4\,\rm{p.u.}$, as shown in Fig.~\ref{fig:powerangle_VA_VSM}. We consider that after the fault has been cleared, the grid voltage recovers to $v_{\rm g}=0.8\,\rm{p.u.}$. From the figure, we can see that the converter initially operates at point ``{a}" at an angle $\delta_{\rm o}$. Following the fault, the nonlinear curve due to current saturation causes the operating point to move from point ``{a}" to point ``{b}". Due to the second-order dynamics of the VSM, the post-fault synchronism can be simply determined using the equal area criterion. It holds that $s_{+}=s_{-}$ at the critical angle $\delta_{\rm c}$ at point ``{c}". Thus, if the fault is cleared before the angle reaches the critical angle $\delta_{\rm c}$, the VSM-controlled GFM converter will maintain synchronization after the fault clearance.

For the droop-controlled GFM converter, we consider two cases with power setpoints $p^{\star}=0.4\,\rm{p.u.}$ and $p^{\star}=0.25\,\rm{p.u.}$ as shown in Fig.~\ref{fig:powerangle_VA_droop}. Due to the first-order dynamics of the droop control, synchronism during the fault period is determined by the existence of stable equilibrium points. For $p^{\star}=0.4\,\rm{p.u.}$, the converter initially operates at point ``{a}" at an angle $\delta_{\rm o}$ and moves to point ``{b}" following the fault. As the power setpoint and the current-saturated curve do not intersect, there is no stable equilibrium, and the angle continues to drift from point ``{b}". However, when the power setpoint is reduced to $p^{\star}=0.25\,\rm{p.u.}$, a stable equilibrium exists during the fault. Thus, the movement is from point ``{a}" to ``{b}", finally synchronizing at ``{c}".

\subsubsection{Virtual Power (VP) Enhancement} The virtual power feedback utilizes the unsaturated current reference $\pha {\hat i}$ and the voltage $\pha{\hat v}$ for power calculation in the droop control. From the equivalent circuit shown in Fig.~\ref{fig:eqtckt}(b), it follows that
\begin{equation} 
 p_{\mathrm{VP}} = \mathrm{Re} \left( \pha{\hat v} \cdot \mathrm{conj}( {\pha{\hat i}} ) \right) = \frac{{v}_{\rm g} \cdot \hat{v}}{{x}_{\rm v} +  \mu \cdot {x}_{\rm g}} \sin\delta. 
 \label{eq:p_VAVP}
\end{equation}
Similar to the virtual admittance control, $\mu$ can be obtained as a function of $\delta$ using an expression akin to \eqref{eq:mu_p_VA}. Therefore, the resulting power expression is not sinusoidal with respect to $\delta$ but is a different nonlinear function. However, since $\mu<1$, the equivalent current-dependent grid impedance decreases, resulting in a larger magnitude of the power-angle curve. Thus, power angle deviations during disturbances are reduced, and the transient stability is improved.

\subsubsection{Cross-Forming (CF) Strategy} In the cross-forming strategy, we use the measured saturated current $\pha {i}_{\mathrm{o}}$ and the voltage $\pha{\hat v}$ for the power calculation as follows \cite{He_crossform_2025}, 
\begin{equation}
\begin{aligned}
    p_{\mathrm{CF}} &= \mathrm{Re} \left({\hat{\pha v}\, \mathrm{conj}(\pha{i}_{\mathrm{o}})}\right) \\
    &= \mathrm{Re} \biggl\{\hat v \angle \hat{\theta}\, \mathrm{conj}\biggl(\frac{\lvert \pha{\hat v}_{\mu} \rvert \angle \hat{\theta} - v_{\mathrm{g}} \angle \theta_{\mathrm{g}}}{j x_{\mathrm{v}} + j x_{\mathrm{g}}} \biggr) \biggr\} \\
    & = \frac{v_{\mathrm{g}} \cdot \hat v }{x_{\mathrm{v}} + x_{\mathrm{g}}} \sin{ \delta}.
\end{aligned}
\label{eq:p_CF}
\end{equation}
Thus, with the cross-forming strategy, the sinusoidal power-angle relationship is maintained even during current saturation.

\begin{figure}
  \begin{center}
  \includegraphics{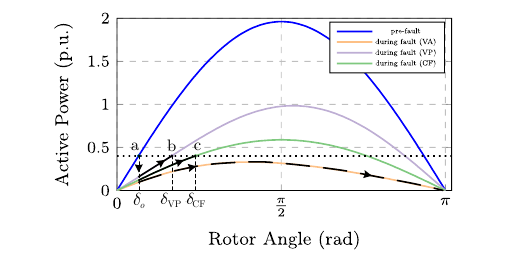}
   \vspace{-0.8cm}
  \caption{Power-angle curves of the three current-limiting strategies considering the droop control operation mode.}
   \vspace{-0.7cm}
    \label{fig:powerangle_all_comparison}
    \end{center}
\end{figure}

Fig.~\ref{fig:powerangle_all_comparison} illustrates an example of the power-angle curves for a droop-controlled GFM converter connected to the grid that operates with the three current-limiting strategies. We can draw the following conclusions. The virtual power enhancement offers superior transient stability due to its ability to dynamically inject power and reduce the power angle. Thus, as shown in Fig.~\ref{fig:powerangle_all_comparison}, the operating point moves from point ``a" to point ``b" following current saturation. The cross-forming strategy prioritizes current limitation and synchronization, but sacrifices dynamic power transfer. Thus, the operating point moves from point ``a" to point ``c" following current saturation as the magnitude of the power-angle curve is lower compared to the virtual power method. The virtual admittance control prioritizes only current protection at the cost of stability margins. Thus, because of the absence of an equilibrium, the angle is observed to drift following current saturation.

\subsection{Simulation Validation}
\begin{figure}
  \begin{center}
  \includegraphics{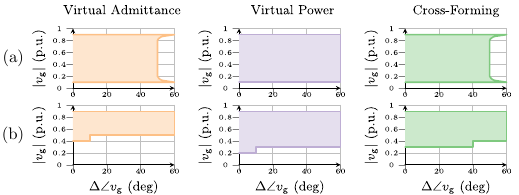}
       \vspace{-0.6cm}
  \caption{{Transient stability results for VSM-controlled, grid-connected GFM converter. Stable region is shaded for (a) $t_{\rm fault}=350\,\rm{ms}$ and (b) $t_{\rm fault}=1\,\rm{s}$.}}
       \vspace{-0.4cm}
    \label{fig:transient_results_VSM}
    \end{center}
\end{figure}

\begin{figure}
  \begin{center}
  \includegraphics{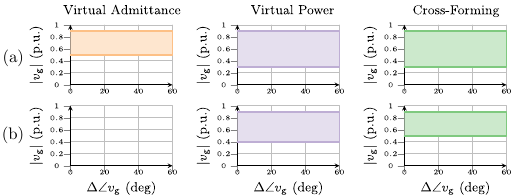}
  \vspace{-0.6cm}
  \caption{{Transient stability results for droop-controlled, grid-connected GFM converter. Stable region is shaded for (a) $p^{\star}=0.5\,\rm{p.u.}$ and (b) $p^{\star}=1\,\rm{p.u.}$.}}
    \label{fig:transient_results_droop}
         \vspace{-0.4cm}
    \end{center}
\end{figure}

We consider the system shown in Fig.~\ref{fig:simulationsystem} to perform detailed EMT time-domain simulations. The base power and current limit are chosen as $S_N=200\,\rm{MVA}$ and $i_{\rm lim}=1.1\,\rm{p.u.}$. The rest of the system parameters are given in Table~\ref{tab:parameters}. Multiple simulations were conducted to analyze the behavior of the different current-limiting strategies under grid faults. The goal is to study the characteristics of transient stability when the residual voltage of the grid is in the range $v_{\rm g}\in[0.1\,\rm{p.u.},\,0.9\,\rm{p.u.}]$ and the grid voltage phase jump is in the range $\Delta\angle{\pha{v}_{\rm g}}\in[0^{\circ},\,60^{\circ}]$. Therefore, simulations are performed for combinations of these events. The stability regions for the VSM-controlled and droop-controlled GFM converter are shown in Figs.~\ref{fig:transient_results_VSM} and~\ref{fig:transient_results_droop}, respectively.

Fig.~\ref{fig:transient_results_VSM} shows the results of the VSM-controlled converter for a fault duration of (a) $t_{\rm fault}=350\,\rm{ms}$ and (b) $t_{\rm fault}=1\,\rm{s}$. For the fault duration $t_{\rm fault}=350\,\rm{ms}$, we can see that the behaviors of cross-forming and virtual admittance strategy control are quite similar. The virtual power enhancement performs better for large phase jumps of \(60^\circ\), due to the higher stability margin. For an increase in the duration of the fault to $t_{\rm fault}=1\,\rm{s}$, the stability region for all current-limiting strategies is reduced. The performance of the virtual admittance strategy significantly deteriorates as the angle $\delta$ increases beyond the critical $\delta_{\rm c}$ for certain faults, therefore, not satisfying the equal area criterion. The virtual power and cross-forming strategies have comparable performance for the increased duration of the fault, however, the former outperforms the latter due to the higher magnitude of its power-angle curve.

Fig.~\ref{fig:transient_results_droop} shows the stability regions of the droop-controlled converter for the power setpoints (a) $p^{\star}=0.5\,\rm{p.u.}$ and (b) $p^{\star}=1\,\rm{p.u.}$. For the lower setpoint $p^{\star}=0.5\,\rm{p.u.}$, the virtual admittance strategy has an inferior performance compared to the virtual power and cross-forming strategies due to the non-existence of stable equilibrium points. This is more prominent for the higher power setpoint $p^{\star}=1\,\rm{p.u.}$ where the virtual admittance strategy is not found to be stable in any case. The virtual power enhancement slightly outperforms the cross-forming strategy due to its higher power margin provision.

\section{Reactive Current Injection Performance}

\subsection{Comparative Analysis}

Reactive current is the current component that lags behind the terminal voltage by $90^\degree$. It is a critical factor in maintaining grid stability, supporting voltage recovery, and ensuring compliance with grid code requirements. In the context of fault conditions, three aspects are analyzed. i) Reactive current magnitude ($i_{Q}$): the magnitude of the reactive current in steady state, ii) start time ($t_{\mathrm{start}}$): the time it takes for $i_{Q}$ to start rising after a fault and iii) rise time ($t_{\mathrm{rise}}$): the time it takes for $i_{Q\rm }$ to rise from 10\% to 90\% of the steady-state value. The magnitude of the reactive current improves the voltage profile, whereas the start and rise times are indicative of the response speed to provide immediate voltage support. The three control strategies differ in their reactive current provision as follows:

\begin{figure}
  \begin{center}
  \includegraphics{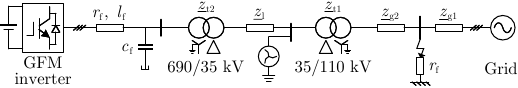}
   \vspace{-0.6cm}
  \caption{Test system used for the simulation validation.}
   \vspace{-0.6cm}
    \label{fig:simulationsystem}
    \end{center}
\end{figure}

\begin{table}
    \centering
    \caption{Test System Parameters}
    \vspace{-0.5cm}\renewcommand\footnoterule{\vspace*{-5pt}}
    \begin{center}
    \scalebox{0.85}{%
    \begin{tabular}{c|l|c}
    \toprule
    \textbf{Symbol} & \textbf{Description} & \textbf{Value} \\ 
    \midrule
    $l_{\rm f}$           & Filter inductance                           & $0.05 \,\mathrm{p.u.}$         \\
    $c_{\rm f }$           & Filter capacitance                          & $0.05 \,\mathrm{p.u.}$         \\
    $r_{\rm f}$           & Filter resistance                           & $0.005 \,\mathrm{p.u.}$        \\
    $T_{J}$             & Inertia time constant                       & $5 \,\mathrm{s}$             \\
    $D$             & Damping coefficient                         & $25$                         \\
    $m_{p}$           & Active power droop gain                   & $0.04$                        \\
    $m_{q}$           & Reactive power droop gain                   & $0.2$                        \\
    $p^*$           & Active power set-point                      & $0.5 \,\mathrm{p.u.}$          \\
    $q^*$           & Reactive power set-point                    & $0.0 \,\mathrm{p.u.}$          \\
    $v^*$           & Voltage set-point                           & $1.0 \,\mathrm{p.u.}$          \\
    $\tau_\mathrm{v}$ & Time constant in $\pha{v}$ feedback        & $0.01 \,\mathrm{s}$          \\
    $\tau_{\mu}$ & Time constant in $\mu$ feedback        & $0.01 \,\mathrm{s}$          \\
    $\pha{z}_{\rm v}$           & Virtual impedance                           & $0.2\,e^{j\pi/2}\,\mathrm{p.u.}$            \\
    $\pha{z}_{\rm{g} 1}$        & Grid impedance 1                            & $0.1005\,e^{j1.4717}\,\mathrm{p.u.}$ \\
    $\pha{z}_{\rm{g} 1}$        & Grid impedance 2                            & $0.03015\,e^{j1.4717}\,\mathrm{p.u.}$ \\
    $\pha{z}_{\rm{t} 1}$        & Transformer impedance 1                            & $0.16016\,e^{j1.538}\,\mathrm{p.u.}$ \\
    $\pha{z}_{\rm{t} 2}$        & Transformer impedance 2                            & $0.06003\,e^{j1.538}\,\mathrm{p.u.}$ \\
    $t_{\rm{fault} }$        & Fault duration                            & $350\,\mathrm{ms}$ \\
    \bottomrule
    \end{tabular}
    }
    \end{center}
    \vspace{-0.5cm}
    \label{tab:parameters}
\end{table}

\subsubsection{Virtual Admittance Control} It regulates the reactive current by operating the reactive power--voltage droop control, the cascaded virtual admittance, and the current saturation.
\subsubsection{Virtual Power Enhancement} The virtual power enhancement injects a larger reactive current compared to the virtual admittance strategy, as a disturbance leads to a smaller change in the power angle causing a smaller active current and a larger reactive current.
\subsubsection{Cross-Forming Strategy} Whenever the cross-forming strategy is activated, $\hat{v}$ is reset to a nominal level, e.g., $1\,\rm{p.u.}$ \cite{He_crossform_2025}, instead of being reduced to a lower level as in the case of reactive power--voltage drop control. Thus, the controller achieves a higher internal voltage, leading to faster and larger reactive current injection. This is particularly true when grid faults are not severe and the current is not saturated.

\subsection{Simulation Validation}

\begin{figure}
  \begin{center}
  \includegraphics{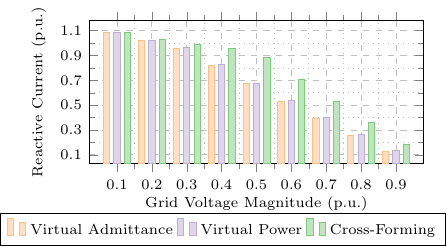}
   \vspace{-0.2cm}
  \caption{{Reactive current magnitude for the VSM-controlled GFM converter under different grid faults for the three current-limiting strategies.}}
   \vspace{-0.6cm}
    \label{fig:iq_VSM}
    \end{center}
\end{figure}

\begin{figure}
  \begin{center}
  \includegraphics{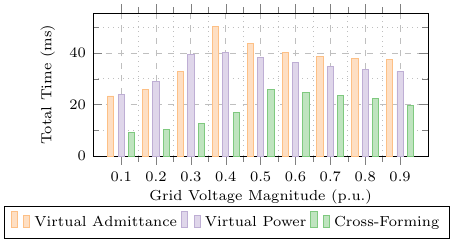}
   \vspace{-0.2cm}
  \caption{{Total time $t_{\rm total}$ for the VSM-controlled GFM converter under different grid faults for the three current-limiting strategies.}}
   \vspace{-0.7cm}
    \label{fig:t_iq_VSM}
    \end{center}
\end{figure}

We consider the test system and parameters described in the previous section to perform detailed EMT simulations to compare the reactive current provision. Again, we consider faults in the grid where the magnitude of the grid voltage is in the range $v_{\rm g}\in[0.1\,\rm{p.u.},\,0.9\,\rm{p.u.}]$. The results for the magnitude and the total reaction time for the VSM-controlled GFM converter are shown in Figs.~\ref{fig:iq_VSM} and~\ref{fig:t_iq_VSM}, respectively. Similar results are obtained for the droop-controlled GFM converter and are therefore not shown in this paper.

Fig.~\ref{fig:iq_VSM} shows the magnitude of the reactive current injected for the VSM-controlled GFM converter under different grid faults. We can conclude that the cross-forming strategy can inject a greater reactive current during faults compared to the other two strategies. This difference is particularly more pronounced for faults of small to medium severity.

Fig.~\ref{fig:t_iq_VSM} shows the total time $t_{\rm total}=t_{\rm rise}+t_{\rm start}$ for the VSM-controlled GFM converter under different faults in the grid. Only the cross-forming strategy has a non-zero start time $(t_{\rm start}\approx0.1\,\rm {ms})$. However, this component has a negligible share in the total time, and the cross-forming strategy outperforms the other two strategies due to its lower rise time. The cross-forming strategy demonstrates enhanced performance across all fault scenarios, delivering superior reactive current injection.

\section{Conclusion}
This paper performs a comparative analysis of three effective current-limiting strategies for GFM converters considering both droop and VSM control. We develop equivalent circuits of the GFM converter for the three strategies to reveal analytical insights into transient stability and reactive current injection. The virtual admittance strategy has a considerably reduced transient stability margin due to the current-dependent equivalent impedance. The cross-forming strategy eliminates the current-dependent nature of the equivalent impedance, thus improving the transient stability. The virtual power enhancement outperforms the former strategies in terms of transient stability, as it is able to considerably increase the stability margin using the unsaturated current reference. In terms of reactive current injection, the virtual power enhancement again outperforms the virtual admittance strategy due to the reduced active power current component. However, the cross-forming strategy outperforms the aforementioned strategies by disabling the voltage-magnitude droop control, thereby maintaining a nominal voltage reference. Finally, we validate these results with detailed EMT simulations considering all possible combinations of grid voltage dips and grid voltage phase jumps. Future work will consider combining the aspects of both the virtual power enhancement and the cross-forming strategy to obtain the aforementioned individual benefits.
\bibliographystyle{IEEEtran}
\bibliography{IEEEabrv,bibliography}

@STRING{IEEE_J_AC         = "{IEEE} Trans. Autom. Control"}

@STRING{IEEE_J_EC         = "{IEEE} Trans. Energy Convers."}

@STRING{IEEE_J_PWRE       = "{IEEE} Trans. Power Electron."}

@STRING{IEEE_J_PWRS       = "{IEEE} Trans. Power Syst."}

@STRING{IEEE_J_STE       = "{IEEE} Trans. Sustain. Energy"}

@STRING{IEEE_J_SG       = "{IEEE} Trans. Smart Grid"}

@STRING{IEEE_J_ESTPE       = "{IEEE} J. Emerg. Sel. Top. Power Electron."}

@STRING{Elsevier_J_EPSR       = "Electr. Power Syst. Res."}

@STRING{Elsevier_J_RSER       = "Renew. Sust. Energ. Rev."}

@STRING{Elsevier_EPES       = "Int. J. Electr. Power Energy Syst."}

@STRING{IEEE_J_IA         = "{IEEE} Trans. Ind. Appl."}

@STRING{IEEE_M_PE         = "{IEEE} Power Energy Mag."}

@article{desai2023saturation,
    author = {Maitraya Avadhut Desai and Xiuqiang He and Linbin Huang and Florian D\"{o}rfler},
    title = {Saturation-Informed Current-Limiting Control for Grid-Forming Converters},
    journal=Elsevier_J_EPSR,
    volume = {234},
    pages = {110746},
    year = {2024},
}

@ARTICLE{huang2019transient,
  author={Huang, Linbin and Xin, Huanhai and Wang, Zhen and Zhang, Leiqi and Wu, Kuayu and Hu, Jiabing},
  journal=IEEE_J_SG, 
  title={Transient Stability Analysis and Control Design of Droop-Controlled Voltage Source Converters Considering Current Limitation}, 
  year={2019},
  volume={10},
  number={1},
  pages={578-591},
  doi={10.1109/TSG.2017.2749259}}

@ARTICLE{li2023transient,
  author={Li, Yujun and others},
  journal=IEEE_J_EC, 
  title={Transient Stability of Power Synchronization Loop based Grid Forming Converter}, 
  year={2023},
  volume={38},
  number={4},
  pages={2843-2859},
  doi={10.1109/TEC.2023.3283396}}

@ARTICLE{fan2022equivalent,
  author={Fan, Bo and Wang, Xiongfei},
  journal=IEEE_J_PWRS, 
  title={Equivalent Circuit Model of Grid-Forming Converters With Circular Current Limiter for Transient Stability Analysis}, 
  year={2022},
  volume={37},
  number={4},
  pages={3141-3144},
  doi={10.1109/TPWRS.2022.3173160}}

@ARTICLE{paquette2015virtual,
  author={Paquette, Andrew D. and Divan, Deepak M.},
  journal=IEEE_J_IA, 
  title={Virtual Impedance Current Limiting for Inverters in Microgrids With Synchronous Generators}, 
  year={2015},
  volume={51},
  number={2},
  pages={1630-1638},
  doi={10.1109/TIA.2014.2345877}}

@article{qoria2020current,
title = {Current limiting algorithms and transient stability analysis of grid-forming {VSCs}},
journal = Elsevier_J_EPSR,
volume = {189},
pages = {106726},
year = {2020},
issn = {0378-7796},
author = {Taoufik Qoria and François Gruson and Fréderic Colas and Xavier Kestelyn and Xavier Guillaud},
}

@article{zhang2023current,
  title={Current-Constrained Power-Angle Characterization Method for Transient Stability Analysis of Grid-Forming Voltage Source Converters},
  author={Zhang, Yu and Zhang, Chen and Yang, Renxin and Molinas, Marta and Cai, Xu},
  journal=IEEE_J_EC,
  year={2023},
  volume={38},
  number={2},
  pages={1338-1349},
  publisher={IEEE}
}

@techreport{gb2024,
  title={The grid code},
  author={{National Grid ESO}},
  institution={National Grid Electricity System Operator},
  year={2024}
}

@techreport{aemo2023voluntary,
    author={{AEMO}},
    title = {Voluntary specification for grid-forming inverters},
    institution = {Australian Energy Market Operator (AEMO)},
    year = {2023}
}

@techreport{entso2024,
    author={{ENTSO-E}},
    title = {GRID FORMING CAPABILITY OF POWER
PARK MODULES},
    institution = {European network of transmission system operators for electricity (ENTSO-E)},
    year = {2024}
}

@article{saffar2023impacts,
  title={Impacts of current limiting on the transient stability of the virtual synchronous generator},
  author={Saffar, Kourosh Gharouni and Driss, Sina and Ajaei, Firouz Badrkhani},
  journal=IEEE_J_PWRE,
  volume={38},
  number={2},
  pages={1509--1521},
  year={2023},
  publisher={IEEE}
}

@article{kkuni2024effects,
  title={Effects of current limit for grid forming converters on transient stability: analysis and solution},
  author={Kkuni, Kanakesh Vatta and Yang, Guangya},
  journal=Elsevier_EPES,
  volume={158},
  pages={109919},
  year={2024},
  publisher={Elsevier}
}

@ARTICLE{wang2023transient,
  author={Wang, Guangyu and Fu, Lijun and Hu, Qi and Liu, Chenruiyang and Ma, Yanhong},
  journal=IEEE_J_STE, 
  title={Transient Synchronization Stability of Grid-Forming Converter During Grid Fault Considering Transient Switched Operation Mode}, 
  year={2023},
  volume={14},
  number={3},
  pages={1504-1515},}

@article{bahrani2024grid,
  title={Grid-Forming Inverter-Based Resource Research Landscape: Understanding the Key Assets for Renewable-Rich Power Systems},
  author={Bahrani, Behrooz and Ravanji, Mohammad Hasan and Kroposki, Benjamin and Ramasubramanian, Deepak and Guillaud, Xavier and Prevost, Thibault and Cutululis, Nicolaos-Antonio},
  journal=IEEE_M_PE,
  volume={22},
  number={2},
  pages={18--29},
  year={2024},
  publisher={IEEE}
}

@ARTICLE{pan2020transient,
  author={Pan, Donghua and Wang, Xiongfei and Liu, Fangcheng and Shi, Rongliang},
  journal=IEEE_J_ESTPE, 
  title={Transient Stability of Voltage-Source Converters With Grid-Forming Control: A Design-Oriented Study}, 
  year={2020},
  volume={8},
  number={2},
  pages={1019-1033},
  doi={10.1109/JESTPE.2019.2946310}}

@article{ghimire2024functional,
  title={Functional Specifications and Testing Requirements of Grid-Forming Offshore Wind Power Plants},
  author={Ghimire, Sulav and Guerreiro, Gabriel MG and Vatta Kkuni, Kanakesh and Guest, Emerson David and Jensen, Kim H{\o}j and Yang, Guangya and Wang, Xiongfei},
  journal={Wind Energy Science Discussions},
  volume={2024},
  pages={1--21},
  year={2024},
  publisher={G{\"o}ttingen, Germany}
}

@article{baeckeland2024overcurrent,
  title={Overcurrent Limiting in Grid-Forming Inverters: A Comprehensive Review and Discussion},
  author={Baeckeland, Nathan and Chatterjee, Debjyoti and Lu, Minghui and Johnson, Brian and Seo, Gab-Su},
  journal=IEEE_J_PWRE,
  year={2024},
  volume={39},
  number={11},
  pages={14493-14517},
  publisher={IEEE}
}

@article{ordono2024current,
  title={Current limiting strategies for grid forming inverters under low voltage ride through},
  author={Ordono, Ander and Sanchez-Ruiz, Alain and Zubiaga, Markel and Asensio, Francisco Javier and Cortajarena, Jose Antonio},
  journal=Elsevier_J_RSER,
  volume={202},
  pages={114657},
  year={2024},
  publisher={Elsevier}
}

@INPROCEEDINGS{Milano-review,
 author={Milano, Federico and Dörfler, Florian and Hug, Gabriela and Hill, David J. and Verbič, Gregor},
 booktitle={Proc. Power Syst. Comput. Conf.},
 title={Foundations and Challenges of Low-Inertia Systems},
 year={2018},
 volume={},
 number={},
 pages={1-25},
 doi={10.23919/PSCC.2018.8450880}}

@article{lasseter2019grid,
  title={Grid-forming inverters: A critical asset for the power grid},
  author={Lasseter, Robert H and Chen, Zhe and Pattabiraman, Dinesh},
  journal=IEEE_J_ESTPE,
  volume={8},
  number={2},
  pages={925--935},
  year={2019},
  publisher={IEEE}
}

@INPROCEEDINGS{virtpower_laba_2023,
  author={Laba, Yorgo and Bruyère, Antoine and Colas, Frédéric and Guillaud, Xavier},
  booktitle={EPE'23 ECCE Europe}, 
  title={Virtual Power-Based Technique for Enhancing the Large Voltage Disturbance Stability of HV Grid-Forming Converters}, 
  year={2023},
  volume={},
  number={},
  pages={1-8},
  doi={10.23919/EPE23ECCEEurope58414.2023.10264486}}

@ARTICLE{He_crossform_2025,
  author={He, Xiuqiang and Desai, Maitraya Avadhut and Huang, Linbin and Dörfler, Florian},
  journal={IEEE Transactions on Power Electronics}, 
  title={Cross-Forming Control and Fault Current Limiting for Grid-Forming Inverters}, 
  year={2025},
  volume={40},
  number={3},
  pages={3980-4007},
  doi={10.1109/TPEL.2024.3500885}}

@ARTICLE{Subotic-dVOC, 
 author={Subotić, Irina and Groß, Dominic and Colombino, Marcello and Dörfler, Florian},
 journal=IEEE_J_AC,   
 title={A {Lyapunov} Framework for Nested Dynamical Systems on Multiple Time Scales With Application to Converter-Based Power Systems},   
 year={2021},  
 volume={66},  
 number={12},  
 pages={5909-5924},  
 doi={10.1109/TAC.2020.3047368}}

\end{document}